\documentclass[12pt]{article}

\usepackage{scicite}

\usepackage{times}

\usepackage{graphicx}
\usepackage{amsmath,bm,enumitem}	
\usepackage[labelfont=bf]{caption}

\usepackage{xcolor}

\newenvironment{sciabstract}{%
\begin{quote} \bf}
{\end{quote}}

\title{Nucleation mechanism of multiple-order parameter ferroelectric domain wall motion in hafnia}

\author
{Songsong Zhou, Andrew M. Rappe$^{\ast}$\\
\\
\normalsize{Department of Chemistry, University of Pennsylvania}\\
\normalsize{Philadelphia, Pennsylvania 19104--6323, USA}\\
\\
\normalsize{$^\ast$To whom correspondence should be addressed; E-mail:  rappe@sas.upenn.edu.}
}

\date{}

\begin{document} 


\baselineskip24pt


\maketitle 

\renewcommand{\figurename}{{\bf{Fig.}}}
\captionsetup{labelfont=bf}


\begin{sciabstract}
Ferroelectric hafnia exhibits promising robust polarization and silicon compatibility for ferroelectric devices. 
Unfortunately, it suffers from difficult polarization switching. Methods to enable easier polarization switching are needed, and the underlying reason for this switching difficulty is not understood. 
Here, we investigated the 180$^\circ$ domain walls of hafnia and their motion through nucleation.
We found that the domains of multiple-order parameter hafnia possess complicated three-dimensional dipole patterns and lead to domain walls of different symmetry.
The most common domain wall type is a complex domain wall involving reversal of both polarization and tetragonality order parameters. 
This domain wall symmetry ensures a good matching of the dipoles perpendicular to the domain wall, which leads to low domain wall energy. However, this ensures a sharp, high energy, charged domain wall on the edges of nuclei that results in difficult nucleation. 
Thus, this domain wall is too stable to move, which explains the switching difficulty of hafnia.
By contrast, another simple domain wall, involving only polarization reversal, has a poor matching of dipoles perpendicular to the domain wall. This leads to higher domain wall energy and ensures a diffusive and low energy charged domain wall that enables easier nucleation.
This simple domain wall is thus not too stable and easier to move.
Our theory advances domain wall nucleation theory from the field of conventional single-order parameter to multiple-order parameters. 
We propose controlling the populations of different domain wall types in hafnia as a way to enable fast polarization switching and lower coercive fields.
\\*
\end{sciabstract}

Ferroelectric materials exhibit spontaneous electric polarization that could be switched by external electric fields.
In the past decade, the interest in ferroelectricity has been heightened by the discovery of ferroelectric hafnia\cite{boscke11p112904,muller11p112901}.
The orthorhombic ferroelectric phase of hafnia is generated through a sequence of phase transitions. The high symmetry cubic phase (C) is first transformed to the tetragonal phase (T) by lowering the temperature\cite{terki08p1484}, and then the tetragonal phase is converted to the orthorhombic phase (O) due to various factors such as surface states, dopants, vacancies, and strain, etc\cite{kelley23p1144,zhou22peadd5953,mittmann19p1900042,materlik18p164101,pevsic16p4601,xu21p826,olsenp12p082905,hyuk14p072901,hoffmann15p072006,shimizu15p112904,ihlefeld23p082901}.  
Unlike conventional perovskite ferroelectric phase transitions involving only polarization as an order parameter, multiple order parameters are involved in fluorite ferroelectric phase transitions\cite{reyes14p140103,qi20p214108}. 
The multiple-order parameter nature of hafnia leads to its unique structure and properties. 
The orthorhombic phase has a structure with an alternating pattern of a polar layer and a non-polar layer within one unit cell, originating from the superposition of equal amplitudes of polar and antipolar order parameters\cite{lee20p1343}.
The strong coupling between these order parameters also offsets the depolarization field effect and hence makes polarization unusually robust even in the thinnest films\cite{zhou22peadd5953}.
Robust ferroelectricity even at the ultra-thin limit, as well as  compatibility with current silicon technology\cite{muller12p4318,boscke11p102903,mikolajick18p340}, make hafnia a promising candidate for various applications including information storage, neuromorphic circuits, and negative differential capacitance transistors\cite{mikolajick18p340, khan15p182, saha21p133701}.
Despite the promising advantages over conventional ferroelectrics, the difficulty in polarization switching and high coercive field are major challenges for practical application of hafnia \cite{zhou15p240,lee19p8929,muller15pN30,wang21p010902}. 
Methods to enable easier switching and to lower the coercive field are urgently needed, and the reason for the difficulty in switching is not completely understood.

The low domain wall mobility is believed to be a key factor for the slow polarization switching and high coercive field. 
An unusually large energy barrier for coherent motion of the entire domain wall plane was found\cite{lee20p1343,ding20p556,zhao22p064104}.
However, such coherent motion, despite its conceptual simplicity, is usually not favorable for domain wall motion. 
Instead, as previously proven in many ferroelectrics, the most common domain wall motion is through a domain wall nucleation mechanism.
In this mechanism, small critical nuclei, with charged domain wall (CDW) as their edges, form adjacent to the domain wall, and these nuclei then grow until the entire domain wall plane is flipped\cite{shin07p881,liu16p360,tybell02p097601,hayashi72p616,miller60p1460}. 
The CDW energy has great impact on the nucleation and switching process, where the critical nucleus represents a balance between the energy gain from dipole alignment with applied electric field and the energy cost of generating more interfacial area.
A higher energy CDW will significantly increase the critical nucleus size, and hence the nucleation energy $\Delta U_{\rm{nuc}}$, which in turn lowers the domain wall velocity (proportional to $\exp[-\Delta U_{\rm{nuc}}/k_BT]$) and increases the coercive field\cite{liu16p360,shin07p881,merz54p690}.

Here, we studied the nucleation on different 180$^\circ$ DWs and the associated CDW properties in hafnia, and we revealed the relationship of their order parameters to their electrostatic profiles and their mobility under electric field.
We found that the multiple-order parameter hafnia domains have complicated three-dimensional dipole patterns that could be understood as 'Lego' blocks, and different types of DWs can be generated by combining these blocks in different ways (i.e., different DW symmetry).  
The different symmetry of DW types distinctly affects the associated CDWs and thus the nucleation behavior.
The most common (lowest-energy) DW type is a complex DW, which is not only a ferroelectric DW with polarization reversal but also a DW of tetragonality order parameter reversal.
This DW symmetry change due to tetragonality reversal leads to abnormal sharp polarization profile preference and low DW energy. 
During DW nucleus formation, this preference forces the associated CDW on the nuclei to adopt a sharp profile that is high in energy, leading to difficult nucleation.
Consequently, this complex DW is too stable to be moved due to its symmetry, which explains the observed difficulty in polarization switching and high coercive field in hafnia.
In contrast, a simple ferroelectric DW without reversing tetragonality shows higher DW energy and diffusive polarization profile preference, similar to conventional perovskite ferroelectrics.
The diffusive preference leads to low energy, diffusive CDW on nuclei, and thus easier nucleation.
This simple DW is not too stable, making it easier to move, which leads to easier polarization switching and lower coercive field.
Our work demonstrated the existence of both complex domain wall and simple ferroelectric domain wall in hafnia due to its multiple-order parameter nature. The two types of domain walls have very different nucleation and domain wall motion behaviors and polarization switching rates, suggesting that controlling the population of different types of DW offers a direction forward to materials and devices that show fast switching and low coercive field.

\section*{Results and Discussions}
To categorize the configurations of different DW types, we first analyzed the modes and order parameters associated with the various phases of hafnia. 
Hafnia is in the cubic fluorite phase (C phase) above 2870 K\cite{terki08p1484}.
The cubic fluorite structure of HfO$_2$ has a edge-sharing lattice of Hf tetrahedral containing an oxygen atom in each cage center, as shown in Fig.1A.
We summarizes this structure as a $2\times2\times2$ Lego block where the 8 oxygen ions are in the center of the  $1\times1\times1$ small cubic boxes. 
Every surface of this block is flat, as oxygen is in the center. 
When the oxygen is displaced along any axis, a Lego post/hole represents outward/inward oxygen displacement.
As temperature is lowered, hafnia is transferred to the tetragonal phase (T phase) by condensing the tetragonal mode (order parameter $T$), where oxygen ions are displaced in parallel along the x-axis with direction alternating based on location in the yz-plane.
The ferroelectric orthorhombic phase (O phase) involves broken symmetry from the T phase by three addtional order parameters. 
The polarization is carried by the polar mode ($P$), involving a uniform z-axis displacement of all oxygens, as well as an additional parallel displacement of oxygen ions along y-axis, with direction alternating based on x-coordinate (see SI Appendix).
The additional antipolar mode ($A$) and non-polar mode ($M$) mainly involve parallel oxygen displacements along the z- and x-axis, with direction alternating based on y- and z-coordinates, respectively (note other minor displacement such as Hf displacement are not listed for simplicity, see SI Appendix for detail). 
These two modes act to cancel/reinforce the z-axis and x-axis displacement induced by $T$ and $P$ in half of the unit cell, forming the alternating polar/nonpolar layers in the O phase.
The structure of any orthorhombic variant can be described by any three independent order parameters from $\{P, A, T, M\}$.
The orthorhombic variant structure is hence described by the sign of three independent order parameters $(P,A,T)$, as $(+++)$, $(++-)$...$(---)$.
Similarly, the cubic and tetragonal phase are also described by vectors $(0,0,0)$ and $(0,0,+)$ or $(00-)$, respectively.

A DW is an interface separating two domains that are related by a symmetry operation to another, which we portray with differently oriented Lego blocks.
Due to the multiple-order parameter nature, various inequivalent DWs could be generated through different symmetry operations. 
Among them, the $(---)/(+++)$ DW, where all parameters reverse their sign across the DW, has the lowest DW energy, and is thus considered to be the most common DW type. 
This is a complex domain wall that is not only a ferroelectric domain wall of reversing polarization but also a domain wall acroos which the tetragonal displacement change sign as well.
The existence of such a complex DW is a result of multiple-order parameter nature of hafnia, which allows a stable intermediate T phase $(0,0,\pm)$ between cubic and orthorhombic phase as well as a stable tetragonal DW $(0,0,-)/(0,0,+)$ of reversed tetragonality.
The two domains (Lego blocks) of this complex DW can be related by a mirror operation noraml to the z-axis, as per Fig.2A. 
Such operation reverses both the polarization direction and the tetragonality (reversing the posts/holes in the top and front surfaces of the Lego blocks) to form the complex DW, but it preserves the displacements along the y-axis (posts and holes in the side surface). 
When combining the two Lego block domains, the posts on the side surface of one block points to the holes of the other block domain, which makes a perfect joining of the two block domains.
Such perfect matching between posts and holes ensures a stable configuration and low DW energy.
In contrast, the two domains forming the simple ferroelectric DW are related by a 180$^\circ$ rotation around the y-axis, which preserves the tetragonality but reverses the posts/holes on the side surfaces, as per Fig.2B. 
Thus, the posts on the side surface of one block domain point not to the holes but to posts of the other block domain, which makes a poor compatibility of the two block domains.
Such a mismatch of posts and holes leads to higher DW energy and less stability compared to the complex DW. 

Besides the different energy and stability, the post/hole matching or mismatching of y-axis dipoles also leads to distinct (z-axis) polarization profile (diffuseness) of the two DWs, which turn out to be important for domain wall nucleation.
The polarization profile of a DW is determined by the interplay between two types of energy contributions. 
The first type is the local bulk energy contribution, the energy cost for the structural distortions that make the local structure deviate from the perfect bulk equilibrium structure.
Such energy contribution would be zero for a perfectly sharp domain wall. 
The second type is the gradient energy contribution, which represents the energy cost due to the discontinuity of order parameters across the DW. 
A more diffusive DW with smaller gradient usually decreases this gradient energy contribution.
Depending on the balance between these two competing contribution, the conventional ferroelectric DW is more or less diffusive.
This also holds true for the simple DW in hafnia.
A diffusive polarization profile (width of 4.5\AA ) is found to minimize the energy for the simple DW, as per red lines in Fig.2C. 
The gradient energy contribution as a function of polarization profile width also shows a negative slope (Fig.2E), confirming that the gradient energy favors a diffusive polarization profile.
This diffusive preference originates from the poor post/hole matching of the simple DW, which forms a high energy head-to-head/tail-to-tail configuration along the y-axis.
As posts/holes on the top and side surfaces are both induced together by the order parameter $P$, a diffusive polarization profile would be most favorable to reduce the mismatching and the corresponding gradient energy. 
In contrast, the complex DW shows an abnormal, perfectly sharp (width of 0\AA) polarization profile, as per black lines in Fig.2C.
The gradient energy vs. width shows a positive slope (Fig.2D), suggesting that a sharper profile would decrease the gradient energy. 
The abnormal sharp polarization profile preference originates from the perfect post/hole matching in the complex DW. 
A diffusive polarization profile will weaken the post/hole matching across the DW, making this Lego ensemble (DW) less stable and leading to higher energy.
In other words, the effect of the good post/hole matching on the side surfaces offsets more than the effect of reversed posts/holes on the top surface, making the sharp polarization profile overall favorable.

Analytically, the distinct profile preferences of these DWs could also be further confirmed and understood through a multiple-order parameter Landau-Ginzburg-Devonshire model. 
The gradient energy of hafnia DW could be described by:
\begin{align*}
\Delta U_{g} = \int_{-\infty}^{+\infty} & \left[ g_{20}{\left(\frac{\partial P}{\partial y} \right)}^2 + g_{02}{\left(\frac{\partial T}{\partial y} \right)}^2 +g_{40}{\left(\frac{\partial P}{\partial y} \right)}^4\right. \\
&\left.+g_{22}{\left(\frac{\partial P}{\partial y} \right)}^2{\left(\frac{\partial T}{\partial y} \right)}^2+g_{04}{\left(\frac{\partial T}{\partial y} \right)}^4\right] dy
\end{align*}
where $g_{mn}$ is the coefficient with the $m,n$ denoting the power of order parameter $P,T$, respectively. Note that $A=P$ and $T=M$ is assumed across the DW to maintain the alternate polar/nonpolar pattern in orthorhombic phase so that the model depends only on $P$ and $T$ (see SI Appendix for detail).
This model perfectly fits the first-principles data, and the leading coefficient of the second-order gradient integral term of $P$ for the complex DW is found to be negative (see SI Appendix Table.S1), which further confirms the intrinsic sharp polarization profile preference.
In contrast, for the simple ferroelectric DW, $T$ does not change sign across the DW so that the corresponding terms can be discarded, leaving only $P$ related terms.
Accordingly, the equation for gradient energy contribution for this DW would have the same form as for the conventional single order parameter ferroelectrics, where the gradient energy increases with decreasing width (Fig.3E).
Consequently, the different symmetry operation to generate the DW (i.e., how the order parameters change across DW or how the Lego blocks are assembled) lead to the distinct profile preferences.

The different polarization profile preferences of the two 180$^\circ$ DW types lead to very different energy for the CDW, which leads to dramatically different nucleation and switching behaviors.
The polarization profile of the CDW on the edge of the nucleus is determined by the competition of three energy contributions: the local energy of nearby bulk, and the gradient energies of the 180$^\circ$ DW and the CDW (Fig.3B and 3E).
Due to the accumulated net interfacial charge, both the complex CDW and simple CDW (single red line) prefer a diffusive polarization profile (marked by red wavy lines) to lower the concentration of charge and hence the gradient energy. 
The local bulk structure (green region) near the nucleus always prefers a sharp polarization profile (green straight lines) to minimize the local bulk energy contribution.
The key factor to determine the polarization profile near the CDW is the two 180$^\circ$ DWs (vertical blue lines circled, which are found to have distinct polarization profile preferences (marked by blue straight lines and wavy lines, respectively).
For the CDW of the nuclei on the complex DW, only the CDW gradient energy contribution prefers a diffusive profile, but both local bulk and 180$^\circ$ DW (due to the good matching of post/hole along the horizontal direction) prefer a sharper profile.
Consequently, the polarization profile of this complex CDW is relatively sharp, as proven by DFT calculation in Fig.3C. 
The sharp polarization profile increases the accumulated net charge density on the CDW, leading to a very large CDW energy of 1.47 J/m$^2$\cite{miller60p1460}. 
This energy is one order of magnitude larger than the CDW energy of typical perovskite such as PbTiO$_3$ (0.12 J/m$^2$). 
Such a high CDW energy would lead to a colossal nucleus. 
For example, under an typical external field of 1.5 MV/cm, the critical nucleus size is estimated to be 20 nm, which is comparable to the typical thickness of hafnia thin-films.
Such large critical nuclei will make the nucleation very difficult and exponentially slow domain wall velocity. 
This may explain the sluggish polarization switching and high coercive field in hafnia thin films. 
By contrast, for the CDW on the nuclei of a simple DW, only the local bulk energy prefers a sharp profile, but both the CDW and the 180$^\circ$ DW (due to the poor matching of posts/holes along horizontal direction) gradient energy prefer a diffusive profile.
Thus, the polarization profile of this CDW is much more diffusive (width of 26\AA) than the complex one (width of 10\AA), as per our calculation in Fig.3F.
The diffusive profile leads to lower net charge density and lower CDW energy of 0.39 J/m$^2$, which is only about one quarter of the complex CDW energy. 
We estimated that this much smaller CDW energy would lead to a critical nucleus size of only 5 nm under the same external field of 1.5 MV/cm. 
The simple ferroelectric DW thus should enable easier nucleation and polarization switching, requiring smaller coercive field.

In summary, the different symmetry operations to generate the complex DW and simple DW lead to their distinct DW energies, nucleation, and switching behaviors.
The symmetry of the complex DW ensures a good matching of post/hole across the DW and thus leads to the low DW energy.
However, the side effect of this good matching is the sharp polarization profile preference, which make the CDW on the edge of nuclei also sharp and thus leads to the difficult nucleation and polarization switching.
In other words, the complex DW is too stable to be moved.
By contrast, the symmetry of the simple DW ensures a poor matching of post/hole and thus has higher DW energy.
However, this poor matching also leads to a diffusive profile preference, which makes the CDW also diffusive and thus has lower CDW energy.
This lower CDW energy enables easier nucleation and polarization switching.
Thus, the simple DW is not too stable an it will be easier to move.

\section*{Conclusion}
We investigated the hafnia DW and its nucleation behavior.
The multiple order parameters hafnia phases are portrayed schematically as Lego blocks. 
Combining the blocks in different ways may leads to complex DW and simple ferroelectric DW of different symmetry, energy, and nucleation behavior.
From a theoretical perspective, our theory highlights the existence of complex DWs and the DW symmetry is the keys to understanding the difficulty in polarization switching and the high coercive field. This approach advances the domain wall motion and nucleation theory from the conventional single-order parameter field to multiple-order parameter materials.  
From a practical perspective, our work suggests that controlling the DW types and populations is a possible way to enable fast polarization switching and lower coercive field.

\section*{Method}
Density functional theory simulations were performed using Vienna Ab initio Simulation Package (VASP)\cite{kresse96p15,kresse96p11169} with a plane-wave basis set and the projector augmented-wave method\cite{blochl94p17953,kresse99p1758}.
The local density approximation (LDA) was used to describe the exchange-correlation energy functional. 
The plane-wave cut-off was set to 500eV.
A $4\times4\times4$, $4\times1\times4$ and $4\times1\times1$ $k$-point mesh was used to sample the Brillouin zone of bulk hafnia, 180$^\circ$ domain wall and charged domain wall, respectively.
All the atomic structures are fully relaxed until the force on each atom is smaller than 0.01 eV/{\AA}.
The figures of atomic structures and phonon mode displacements are generated with the VESTA code\cite{Momma21p1272}.
The 180$^\circ$ DWs were simulated using a $1\times8\times1$ supercell containing two 180$^\circ$ DWs. 
The 180$^\circ$ DW energy was calculated by subtracting the reference bulk energy from the supercell total energy.
The CDW was simulated using a tilted supercell where both 180$^\circ$ DW and CDW are present. 
The supercell axis is defined by $\vec{a}'=\vec{a}, \vec{b}'=m*\vec{b}, \vec{c}'=n*\vec{c}-\vec{b}$, where the $\vec{a}',\vec{b}',\vec{c}'$ and $\vec{a},\vec{b},\vec{c}$ are the lattice vector of the the supercell and unitcell of orthorhombic hafnia, respectively.
The supercell with $m=6,n=8$ was used to simulate the CDW.
The CDW energy were calculated by subtracting the corresponding 180$^\circ$ domain wall energy and the bulk energy from the supercell total energy. 
To estimate the nuclei size, the nuclei on the 180$^\circ$ DW is assumed to be rectangular, where the aspect ratio is determined by the ratio of the CDW energy and the $(100)$ DW energy.
The energy to form a given size nuclei is estimated by subtracting the edge energy cost from the energy gain of the polarization flipping. 
The critical nuclei size and the nucleation energy is then estimated by determining the saddle point of the energy vs. size curve. 
The polarization profile was obtained by extracting the polar mode amplitude $P$ from each unit cell within the fully relaxed supercells. 

{\bf{Data, Materials, and Software Availability.}}
All study data are included in the article and/or SI Appendix.
\section*{Acknowledgement}
This research is supported as part of the center for 3D Ferroelectric Microelectronics (3DFeM), an Energy Frontier Research Center funded by the U.S. Department of Energy (DOE), Office of Science, Basic Energy Sciences under award no. DE-SC0021118. Computational support was provided by the National Energy Research Scientific Computing Center (NERSC), a U.S. Department of Energy, Office of Science User Facility located at Lawrence Berkeley National Laboratory, operated under Contract No. DE-AC02-05CH11231.

\begin{figure*}
\begin{center}
  \includegraphics[scale=0.45]{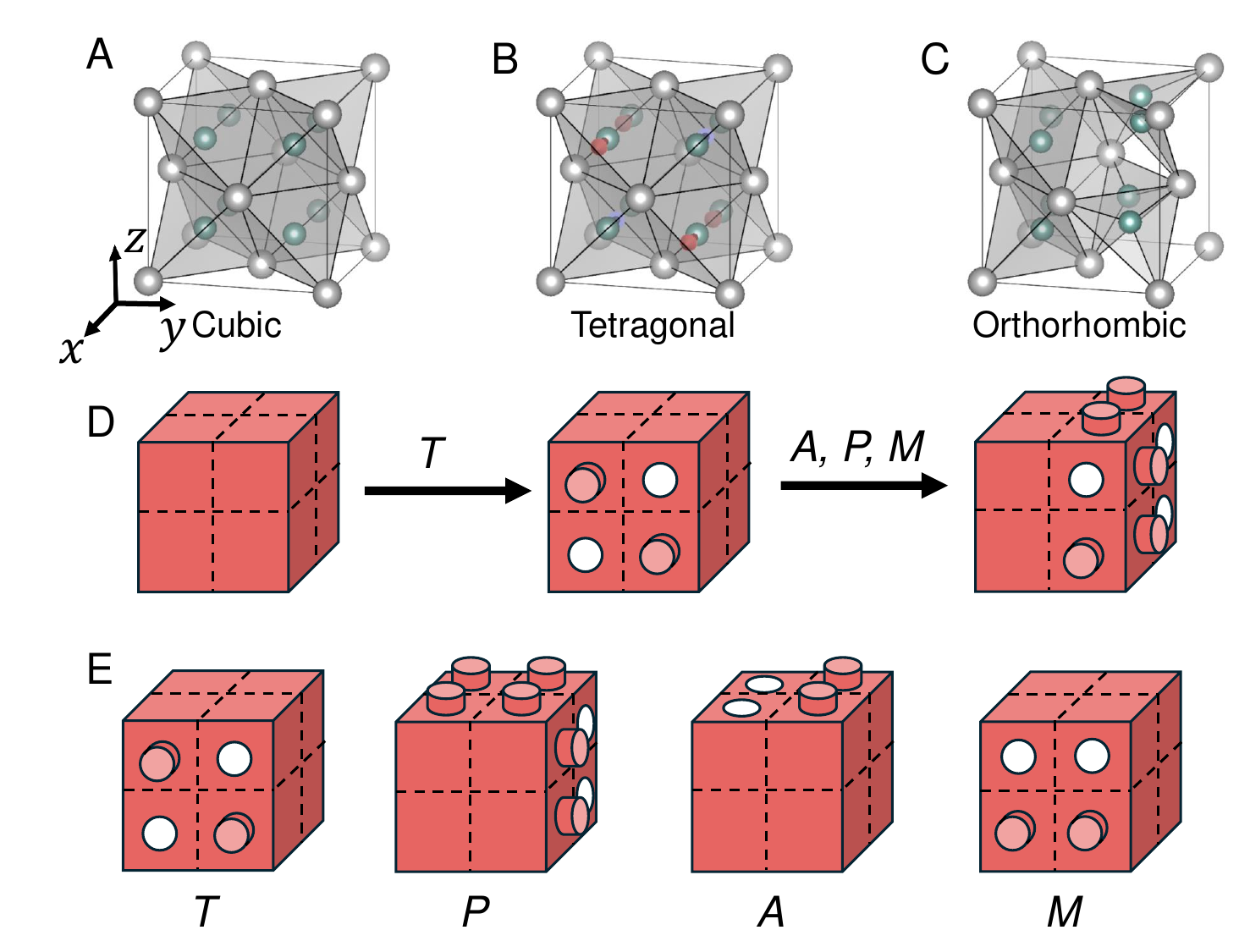}
  \caption{\textbf{Phases and order parameters of HfO$_2$.} Three phases involved in the formation of ferroelectric HfO$_2$. The high symmetry cubic phase (Fig.1A) is of fluorite structure with an oxygen atom (green balls) in the center of each tetrahedral cage of Hf (grey balls). Such structure could be described by a $2\times2\times2$ 'Lego' block, as per the left in Fig.2D. Every surface of the block is flat, as oxygen is undisplaced in the center of each $1\times1\times1$ block. When the oxygen is displaced toward/away from any surface, a post/hole will be generated on the corresponding surface. As temperature is lowered, hafnia undergoes a phase transition to the tetragonal phase (Fig.1B) by condensing the tetragonal mode (order parameter $T$), which involves parallel oxygen displacement along the x-axis with direction alternating based on location in yz-plane, as per Fig.1E. Note that for clarity only the top, front and right surfaces of the block are plotted. Just like Lego blocks, any post/hole on one surface will have hole/post on the opposite surface. The ferroelectric orthorhombic phase (Fig.1C) is generated from the tetragonal phase by condensing three more order parameters, the polar mode $P$, antipolar mode $A$, and non-polar mode $M$. The polar mode $P$ involves uniform oxygen displacement along z-axis as well as the parallel oxygen displacement along y-axis with direction alternating based on x-coordinate. The $A$ and $M$ order parameter involve the parallel oxygen displacement along z- and x-axis, with direction alternating based on y- and z-coordinate, respectively.}
  \label{figure1}
\end{center}
\end{figure*}

\begin{figure*}
\begin{center}
  \includegraphics[scale=0.3]{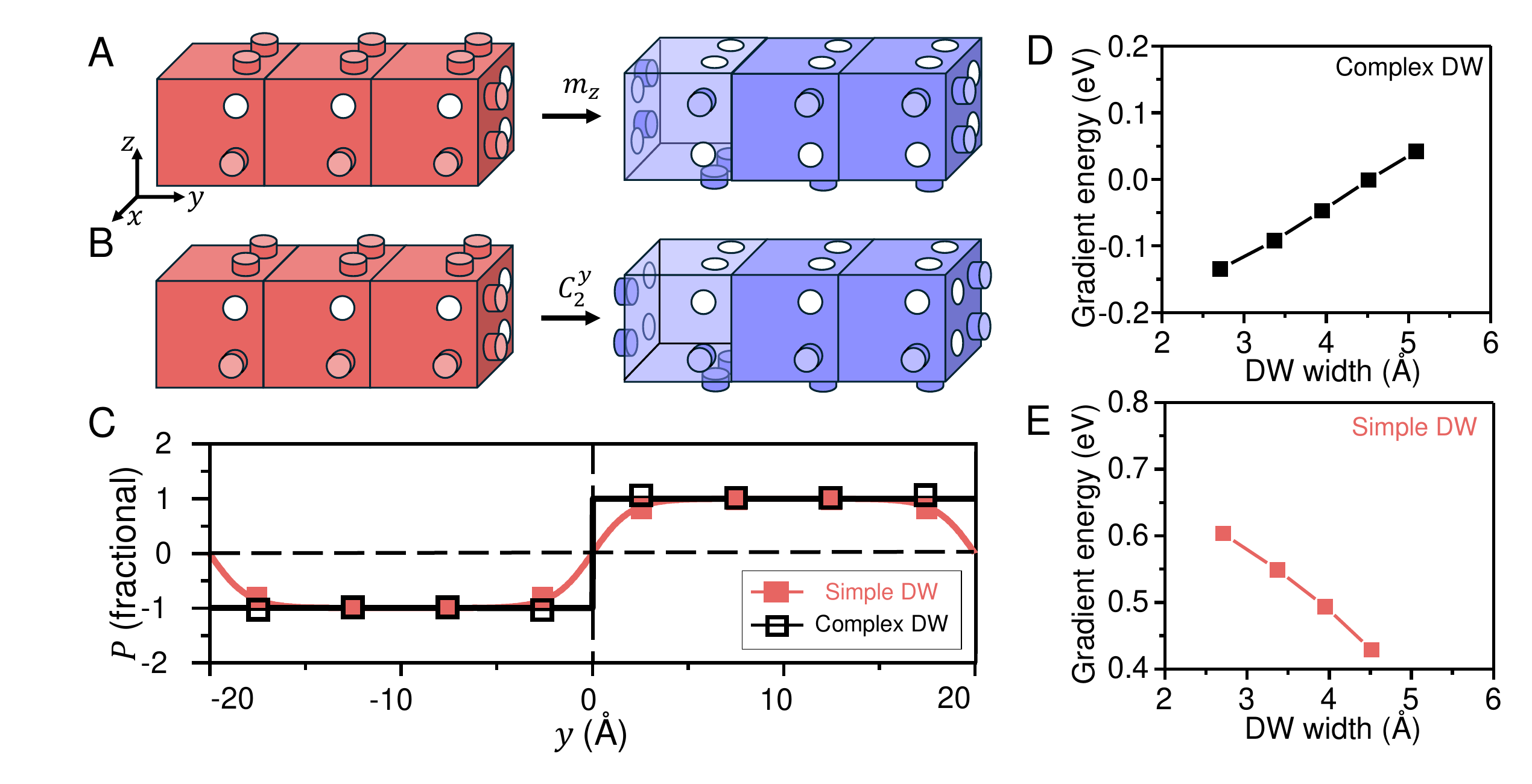}
  \caption{\textbf{Symmetry, energy, and polarization profile of 180$^\circ$ DW}. (A) The two domains (Lego blocks) in complex DW are related by a $m_z$ mirror operation so that the posts/holes on the side surfaces are preserved, which forms a perfect matching of posts/holes across the DW and leads to low DW energy. For the simple DW (B), the two domains are related by a $C_2^y$ rotation operation, which reverses the posts/holes on the side surface so that the posts/holes across the DW are mismatched, which leads to higher DW energy. (C) The different matching of posts/holes also leads to different polarization profile preferences for the two DWs, as shown by the diffusive profile of the simple DW (red) and the abnormal sharp profile of the complex DW (black). (D) the sharp profile preference of complex DW is confirmed by calculating the gradient energy as a function of polarization profile width, where a positive slope suggest a lower gradient energy at smaller width (sharper profile). In contrast, the negative slope confirms the diffusive profile preference of simple DW (E). }
  \label{figure2}
\end{center}
\end{figure*}

\begin{figure}
\begin{center}
  \includegraphics[scale=0.3]{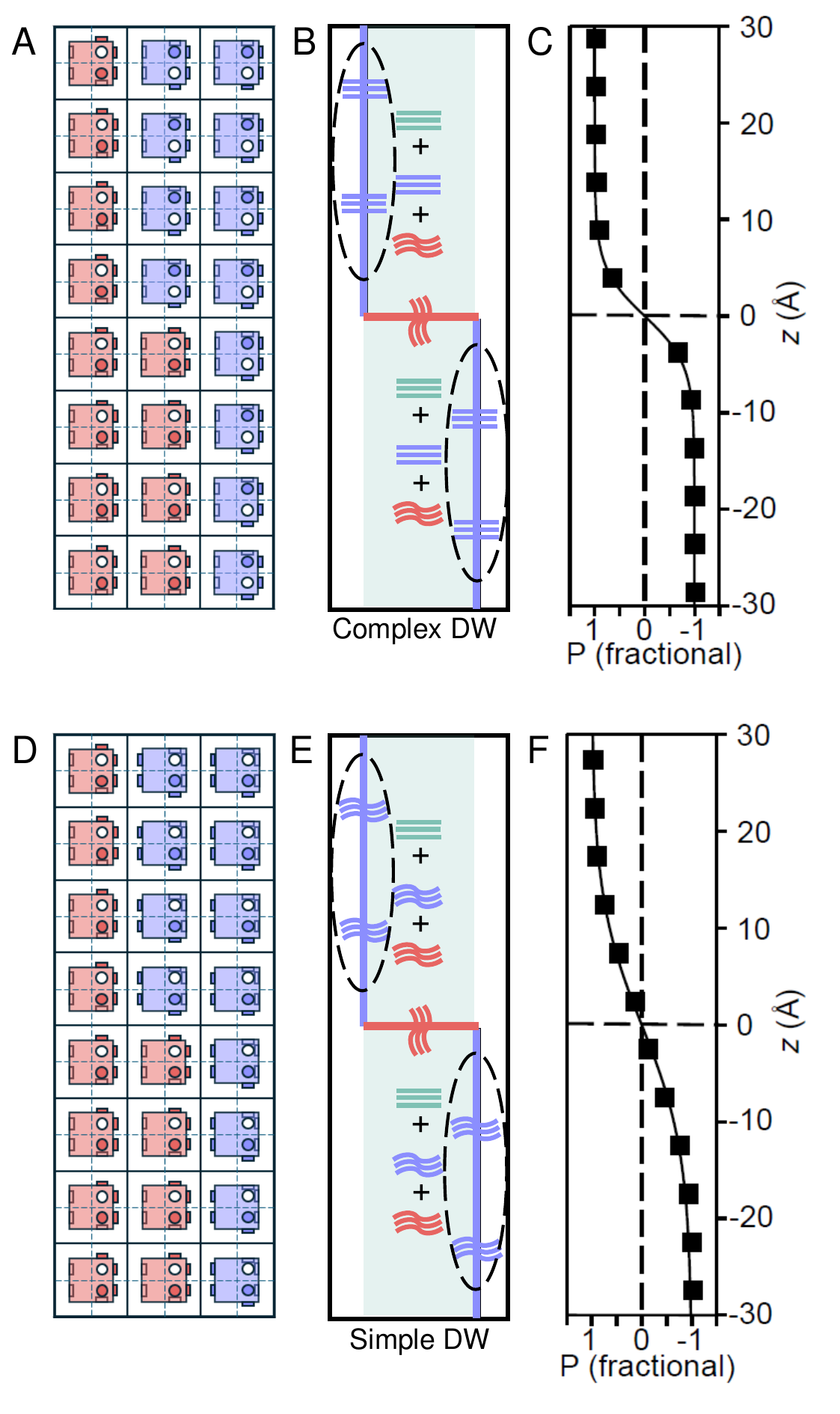}
  \caption{\textbf{The structure and polarization profile of charged domain wall on the edge of nuclei.} (A)(D) the 'Lego' configuration of CDWs on the edge of nuclei for (A) complex DW and (D) simple DW, respectively. Note that in the schematic Lego plots, only the side posts/holes near front face are plotted for clarity. (B)(E) the schematic illustration of the effect of the 180$^\circ$ DW's polarization profile preference on the charged domain wall (CDW) on the edge of nuclei. The polarization profile of the CDW is the result of three competing energies: the energy of local bulk (green region), the gradient energy of the CDW (red horizontal line), and the gradient energy of the 180$^\circ$ DW (blue vertical line). The straight lines and wavy lines denote the preferences for sharp and diffusive profile, respectively. In the complex DW (B), due to the good matching of posts/holes in the horizontal direction as per (A), both the 180$^\circ$ DW and the local bulk prefer a sharp profile, which makes the CDW polarization profile to be relatively sharp, as per (C). The sharp profile leads to high CDW energy and difficult nucleation. In the simple DW (E), due to the poor matching of posts/holes on the horizontal direction as per (D), both the 180$^\circ$ DW and the CDW prefer a diffusive profile, which makes the CDW profile more diffusive, as per (F). This diffusive profile leads to lower CDW energy and easier nucleation. }
  \label{figure3}
\end{center}
\end{figure}

\pagebreak
\bibliography{manuscript1}

\bibliographystyle{Science}

\end{document}